\documentclass[prd,twocolumn,showpacs,preprintnumbers,nofootinbib,amsmath,amssymb]{revtex4}

\usepackage{graphicx}
\usepackage{dcolumn}
\usepackage{bm}
\usepackage{amsmath}

\newcommand{\beq}{\begin{eqnarray}}
\newcommand{\eeq}{\end{eqnarray}}

\newcommand{\ie}{{i.e. }}

\newcommand{\real}{{\sf I}\kern-.12em{\sf R}}
\newcommand{\comp}{{\sf I}\kern-.50em{\sf C}}
\newcommand{\unity}{{\sf I}\kern-.54em{\sf 1}}

\def\spose#1{\hbox to 0pt{#1\hss}}
\def\ltapprox{\mathrel{\spose{\lower 3pt\hbox{$\mathchar"218$}}
 \raise 2.0pt\hbox{$\mathchar"13C$}}}

\begin{document}

\preprint{GEF-TH 18/07}

\title{Imaginary chemical potentials and the phase
of the fermionic determinant}
\author{Simone Conradi and Massimo D'Elia}
\affiliation{Dipartimento di Fisica dell'Universit\`a di Genova and INFN, Sezione di Genova, Via Dodecaneso 33, I-16146 Genova, Italy}

\date{\today}

\begin{abstract}
A numerical technique is proposed for an efficient numerical
determination of the average phase factor of the fermionic determinant
continued to imaginary values of the chemical potential.
The method is tested in QCD with eight flavors of dynamical
staggered fermions. A direct check of the validity of 
analytic continuation is made on small lattices  and a study 
of the scaling with the lattice volume is performed.
\end{abstract}

\pacs{11.15.Ha, 12.38 Gc, 12.38.Aw}

\maketitle

\section{Introduction}

Lattice QCD simulations in presence of a finite density of baryonic
matter are hindered by the well known sign problem. Consider
for instance the QCD partition function
\beq
Z(\mu,\mu) &\equiv&
\int \mathcal{D}U e^{-S_{G}[U]} (\det M[U,\mu])^2 \nonumber \\
&=& 
\int \mathcal{D}U e^{-S_{G}[U]} |\det M[U,\mu]|^2 e^{i 2 \theta} 
\; ,
\label{Zfinbar}
\eeq
describing two flavors of quarks (or eight flavors in the case
of staggered flavors) which are given an equal chemical potential
$\mu$: the determinant of the fermionic matrix $M$ 
is in general complex ($\theta \neq 0$) 
for $\mu \neq 0$ and Monte Carlo simulations are not feasible.
Various possibilities
have been explored to circumvent the problem, like 
reweighting techniques~\cite{glasgow,fodor,density},
the use of an imaginary chemical potential either 
for analytic continuation~\cite{muim,immu_dl,azcoiti,chen,giudice,cea,sqgp} 
or for reconstructing the canonical
partition function~\cite{rw,cano1,cano2}, Taylor expansion
techniques~\cite{taylor1,taylor2} and
non-relativistic expansions~\cite{hmass1,hmass2,hmass3}.

The same is not true in the case of a finite isospin density,
\ie when quarks are given opposite chemical potentials.
Indeed, due to the property $\det M[U,-\mu] = \det M[U,\mu]^*$,
the partition function
\beq
Z(\mu, -\mu) =
\int \mathcal{D}U e^{-S_{G}[U]} |\det M[U,\mu]|^2 
\; 
\label{Zfiniso}
\eeq
has a positive measure. That is also known as phase quenched QCD.
The average value of the phase factor of the
fermionic determinant, 
$\langle e^{i 2 \theta} \rangle_{(\mu, -\mu)}$, where the index 
indicates the partition function the expectation value refers to,
gives a direct measurement of the severeness of the sign problem.
$\langle e^{i 2 \theta} \rangle \sim 0$ will signal the stage
at which the complex nature of the determinant will imply
a significant difference between finite baryonic density and 
finite isospin density, as well a poor reliability of reweighting
techniques (see Ref.~\cite{splitt0} and references therein).

It clearly follows from Eqs.~(\ref{Zfinbar}) and (\ref{Zfiniso})
that the average phase factor is the expectation value of the ratio
of two determinants; it can also be expressed as the ratio
of two partition functions:
\beq
\langle e^{i 2 \theta} \rangle_\mu  &\equiv& 
\left\langle {\det M (\mu) \over \det M (-\mu)}
\right\rangle_{(\mu,-\mu)} = {Z(\mu,\mu) \over Z(\mu,-\mu)} \, .
\label{phase}
\eeq
Its direct numerical computation reveals a difficult numerical
task as the lattice volume $V$ increases, since it involves
the numerical evaluation of fermionic determinants.

It has been proposed recently~\cite{splitt1,splitt2}  to study
the analytic continuation of the average phase factor
to imaginary values of the chemical potential
\beq
\hspace{-10pt}
\langle e^{i 2 \theta} \rangle_{i \mu}  &\equiv& 
\left\langle {\det M (i \mu) \over \det M (- i \mu)}
\right\rangle_{(i\mu,-i\mu)} = {Z(i \mu,i \mu) \over Z(i \mu,- i \mu)} 
\nonumber \\
\hspace{-10pt}
&=& {\int \mathcal{D}U e^{-S_{G}[U]} \det M[U,i \mu] \det M[U, i \mu] 
 \over {\int \mathcal{D}U e^{-S_{G}[U]} \det M[U, i \mu] \det M[U,
     - i \mu]}} \, 
\label{phase2}
\eeq
where $Z(i \mu,i \mu)$ and 
$Z(i \mu,-i \mu)$ are the analytic continuation of the partition
functions at finite baryonic and isospin chemical potentials
respectively, which are both suitable for numerical simulations
since $\det M[U,i \mu]$ is always real.
Numerical difficulties however are present also in this case:
the observable to be averaged is still expressed in terms of
fermionic determinants. Moreover in principle sampling problems
deriving from a bad overlap between the two statistical distributions
described by $Z(i \mu,i \mu)$ and 
$Z(i \mu,-i \mu)$ may arise.
In Ref.~\cite{splitt2} the fermionic determinant has been estimated on
the basis of the lowest lying eigenvalues
of the fermionic matrix.

In the present paper we propose a new technique which, 
making use of numerical strategies developed in different
contexts, permits an exact evaluation of the average phase
factor with a reasonable scaling of the required CPU time
as the lattice volume is increased. In doing this we will
fully exploit the possibility of performing numerical 
simulations of the partition function $Z(i \mu_1, i \mu_2)$ 
for generic values of $\mu_1$ and $\mu_2$.

In Section~\ref{method} we illustrate two different
possible methods, which are then numerically tested and compared
in Section~\ref{results} for the theory
with 8 staggered flavors.

\section{The method}
\label{method}

The evaluation of the average phase factor, expressed like in
Eq.~(\ref{phase}) or Eq.~(\ref{phase2}) as the ratio of two
different partition functions, resembles similar problems
which are encountered in quite different contexts, like
the evaluation of disorder parameters in statistical models and in 
lattice gauge theories. Explicit examples are given by monopole
disorder parameters or by the 't Hooft loop, which enters in various
studies about color confinement. 
The major problem in those cases is the small overlap 
between the statistical distributions corresponding to
two different partition functions, resulting in a poor
sampling efficiency.
Powerful techniques have been developed in both
cases, consisting in either determining derivatives of the 
disorder parameters, from which the ratio of partition functions can
then be reconstructed after integration~\cite{deldeb, sch, vort}, 
or in making use of various reweighting
techniques, like that of rewriting the original ratio 
in terms of intermediate ratios which are more easily 
evaluable~\cite{deforc,helicity,muu1}.

In the present case the major difficulty derives from a direct
computation of the observables, which is expressed in terms of 
fermionic determinants, but sampling problems may in principle
worsen the situation also in this case, especially in the large
volume limit.
In the following we will describe the application of both kind
of techniques described above to the present case, and try to understand
by numerical simulations which of them is best suited in this
context.

We describe at first how to reconstruct $\langle e^{i 2 \theta}
\rangle$ in terms of derivatives. Consider the modified ratio 
\beq
R_\mu(\nu) = 
{Z(i \mu, i \nu) \over Z(i \mu, - i \mu)}
\label{ratio1}
\eeq 
where
\beq
Z(i \mu, i \nu) \equiv  
\int \mathcal{D}U e^{-S_{G}[U]} \det M[U,i \mu] \det M[U, i \nu] \, .
\label{partfun1}
\eeq
It is clear that $R_\mu(-\mu) = 1$, while $R_\mu(\mu)$ is the original
ratio. It can be easily verified
that
\beq
\rho(\nu) &\equiv&
\frac{d}{d \nu} \ln R_\mu(\nu) = 
\frac{d}{d \nu} \ln  Z(i \mu, i \nu) \nonumber \\
&=& 
\left\langle 
i\ {\rm Tr} \left( M^{-1}( i \nu) \frac{d}{d (i \nu)} M(i \nu) \right) 
\right\rangle_{(i \mu, i \nu)} \, .
\label{rhodef}
\eeq
Last quantity is nothing but $i$ times the average number of
quarks coupled to the chemical
potential $i \nu$: it is purely imaginary for symmetry
reasons, hence $\rho(\nu)$ is real,
and can be computed using a noisy unbiased estimator.
The average phase factor can then be obtained by integration
\beq
\langle e^{i 2 \theta} \rangle_{i \mu} = 
\exp\left( \int_{-\mu}^{\mu} \rho(\nu) d \nu \right) \, 
\label{1stmet}
\eeq
and no quark determinant must be explicitly computed.

In practice, the derivative $\rho(\nu)$ will be computed for 
a discrete set of values of $\nu$ and then integrated numerically.
The precision attained for  
$\langle e^{i 2 \theta} \rangle_{i \mu}$
will depend both on the statistical errors of the single determinations
and on the systematic uncertainty linked to numerical integration;
the last can be estimated for instance by varying the chosen interpolation 
procedure. In principle it is also possible to determine further
derivatives of $\rho$ in order to improve the integration accuracy.

As a different method we consider rewriting 
$\langle e^{i 2 \theta} \rangle_{i \mu}$
as the product of $N$ intermediate ratios:
\beq
\langle e^{i 2 \theta} \rangle_{i \mu} =  
{Z(i \mu, i \mu) \over Z(i \mu, - i \mu)} = 
\frac{Z_N}{Z_{N-1}} 
\, \frac{Z_{N-1}}{Z_{N-2}}
\, \dots \, \frac{Z_1}{Z_{0}} 
\equiv
\prod_{k = 1}^{N} r_k
\label{2ndmet}
\eeq
where $Z_N \equiv Z(i \mu, i \mu)$, 
$Z_0 \equiv Z(i \mu, - i \mu)$ while 
\beq
\hspace{-3pt}
Z_k \hspace{-2pt}
\equiv \hspace{-3pt}
\int \hspace{-3pt} \mathcal{D}U 
e^{-S_{G}[U]} \hspace{-1pt} \det M[U,i \mu] \hspace{-1pt} 
\det M[U, i (- \mu + k
\delta \nu)]
\eeq
with $\delta \nu = 2 \mu / N$.
The idea is to compute each single ratio $r_k$ by a different Monte Carlo
simulation. Apart from the increased overlap among 
each couple of partition functions, 
an improvement comes also from the simpler form in which
the observable appearing in
each ratio $r_k$ can be rewritten, 
for large enough $N$.
Indeed we have:
\beq
r_k &=& \left\langle
{\det M (i ( \nu + \delta \nu)) / \det M (i \nu)}
\right\rangle_{(i \mu, i \nu)} \nonumber \\ &=&  
\left\langle
\exp \left( {\rm Tr} \ln A ( \nu, \delta \nu)
\right) 
\right\rangle_{(i \mu, i \nu)}
\label{robs}
\eeq
where $\nu = -\mu + (k - 1) \delta \nu$ and
\beq
A [U, \nu, \delta \nu] \equiv {M[U,i \nu]}^{-1} M [U,i ( \nu + \delta \nu)] 
\, .
\label{Amatrix}
\eeq

If $\delta \nu$ is small, the matrix 
$A [U, \nu, \delta \nu]$ is very close to the identity
matrix ${\rm Id}$ for each configuration $U$. 
We can therefore expand the logarithm in 
Eq.~(\ref{robs}) thus rewriting the following approximate
expression for $r_k$:
\beq
\hspace{-10pt} r_k &\simeq& 
\left\langle
\exp \left( {\rm Tr} (A - {\rm Id}) -\frac{1}{2} {\rm Tr} (A - {\rm Id})^2 + 
\dots
\right) 
\right\rangle
\label{logexp}
\eeq

Each trace in the exponential can be evaluted by a noisy
estimator as follows:
\beq
\hspace{-10pt}
{\rm Tr} (A[U] - {\rm Id})^n 
&\simeq&
\frac{1}{K} \sum_{j = 1}^K \eta^{(j)\, \dagger}
(A[U] - {\rm Id})^n
\eta^{(j)} \, 
\label{noise}
\eeq
where $\eta^{(j)}$ is a random vector satisfying
$\langle \eta^{(j)\, \dagger}_{i_1} \eta^{(j)}_{i_2} \rangle_\eta  = \delta_{i_1,i_2}$.
The computation of each noise estimate in  
Eq.~(\ref{noise}) can be made faster if, when applying the matrix
$A[U] = {M[U,i \nu]}^{-1} M [U,i ( \nu + \delta \nu)]$ 
to the vector $\eta^{(j)}$ (or to $
(A[U] - {\rm Id})^{n-1}
\eta^{(j)}$
at higher orders), $\eta^{(j)}$ itself is taken
as a starting tentative solution for the inverter 
giving ${M[U,i \nu]}^{-1} 
(M [U,i ( \nu + \delta \nu)] \eta^{(j)})$: the guess is
better and better as $\delta \nu \to 0$.

\begin{figure}[t!]
\includegraphics*[width=1.0\columnwidth]{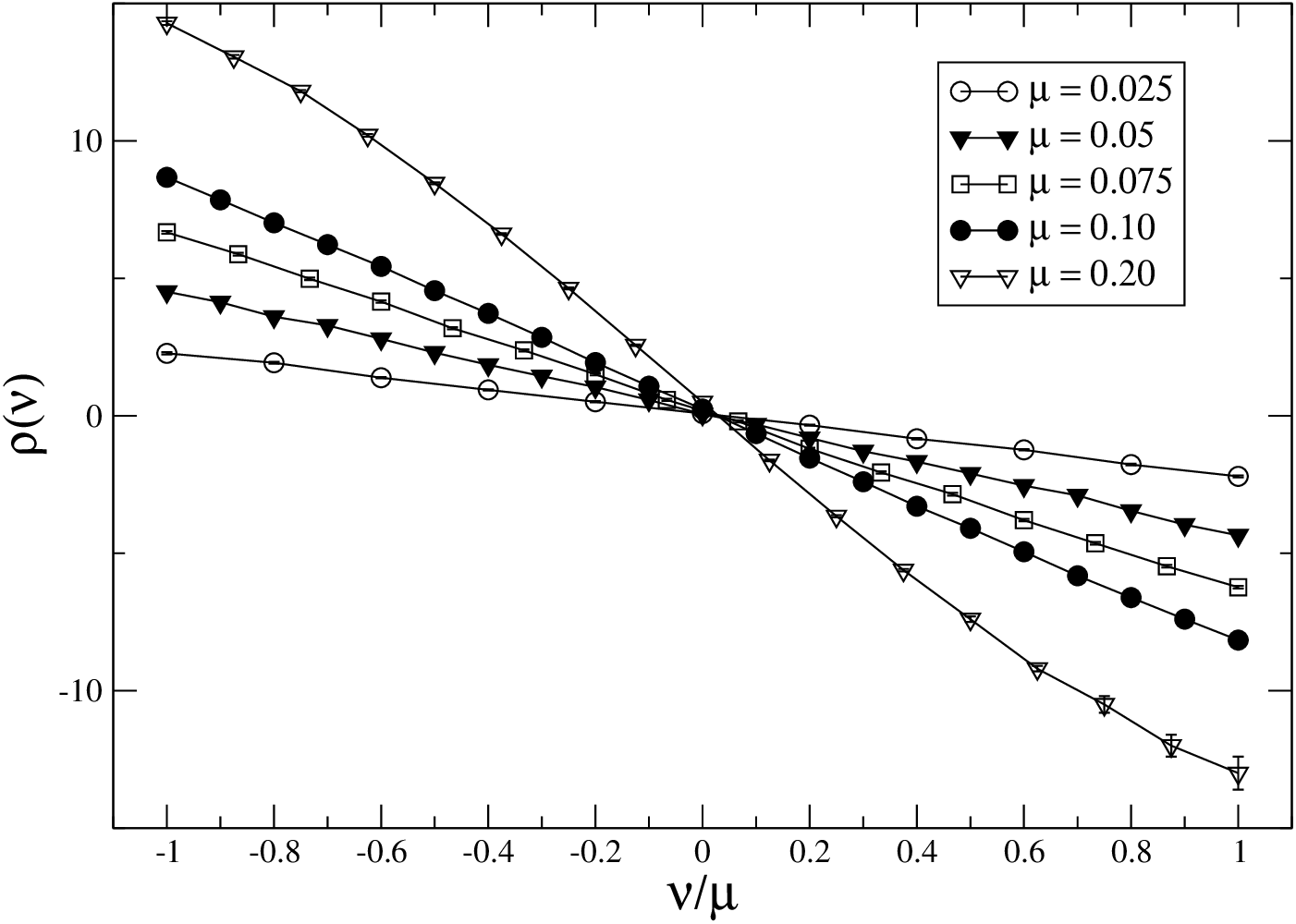}
\vspace{-0.cm}
\caption{
$\rho(\nu)$ for various values of $\mu$ at $\beta = 4.8$ 
and $L_s = 4$.
}
\label{fig1} 
\vspace{-0.cm}
\end{figure}

\begin{figure}[b!]
\includegraphics*[width=1.0\columnwidth]{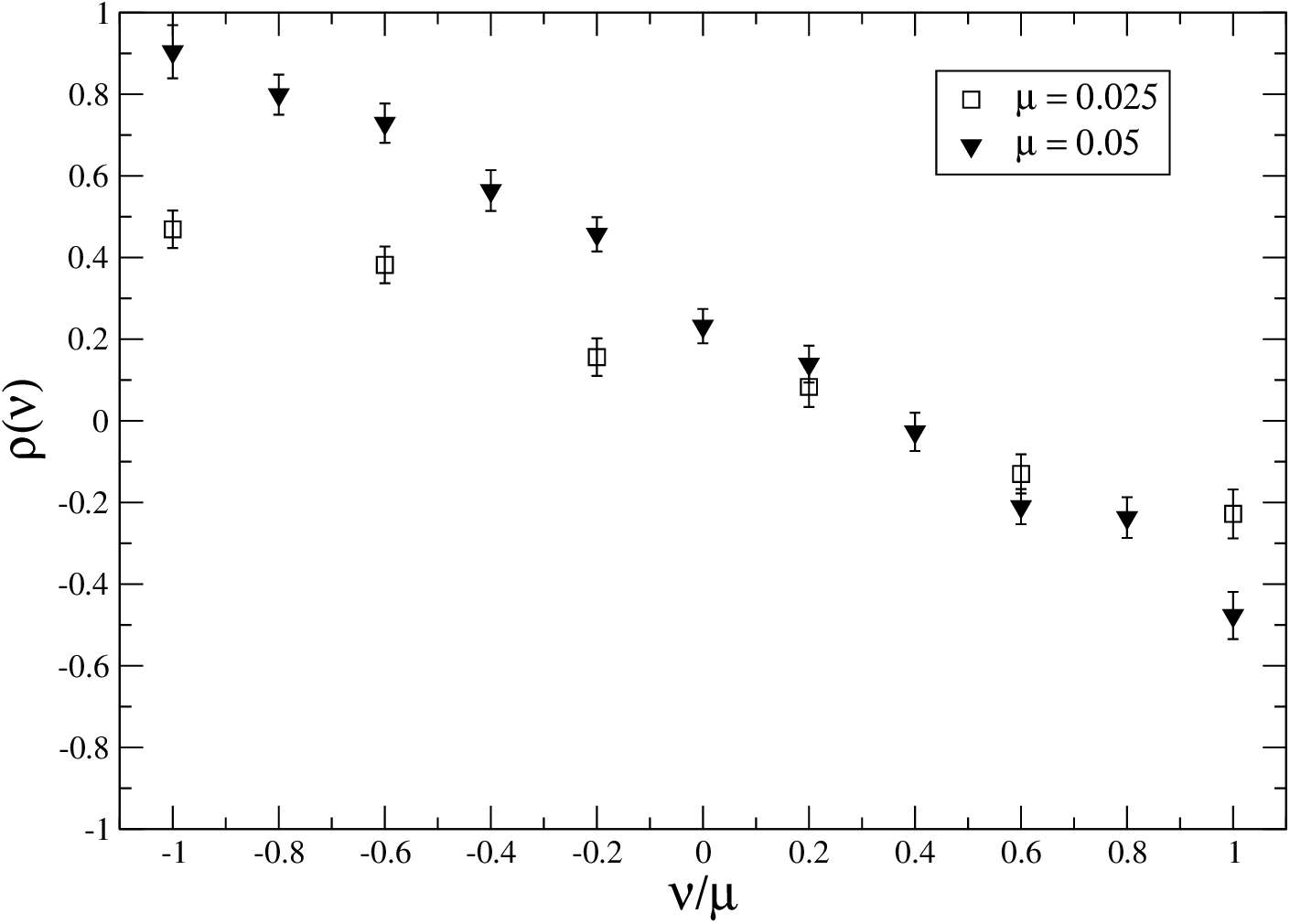}
\vspace{-0.cm}
\caption{
$\rho(\nu)$ for various values of $\mu$ at $\beta = 4.6$ 
and $L_s = 4$.
}
\label{fig2} 
\vspace{-0.cm}
\end{figure}

\begin{figure}[t!]
\includegraphics*[width=1.0\columnwidth]{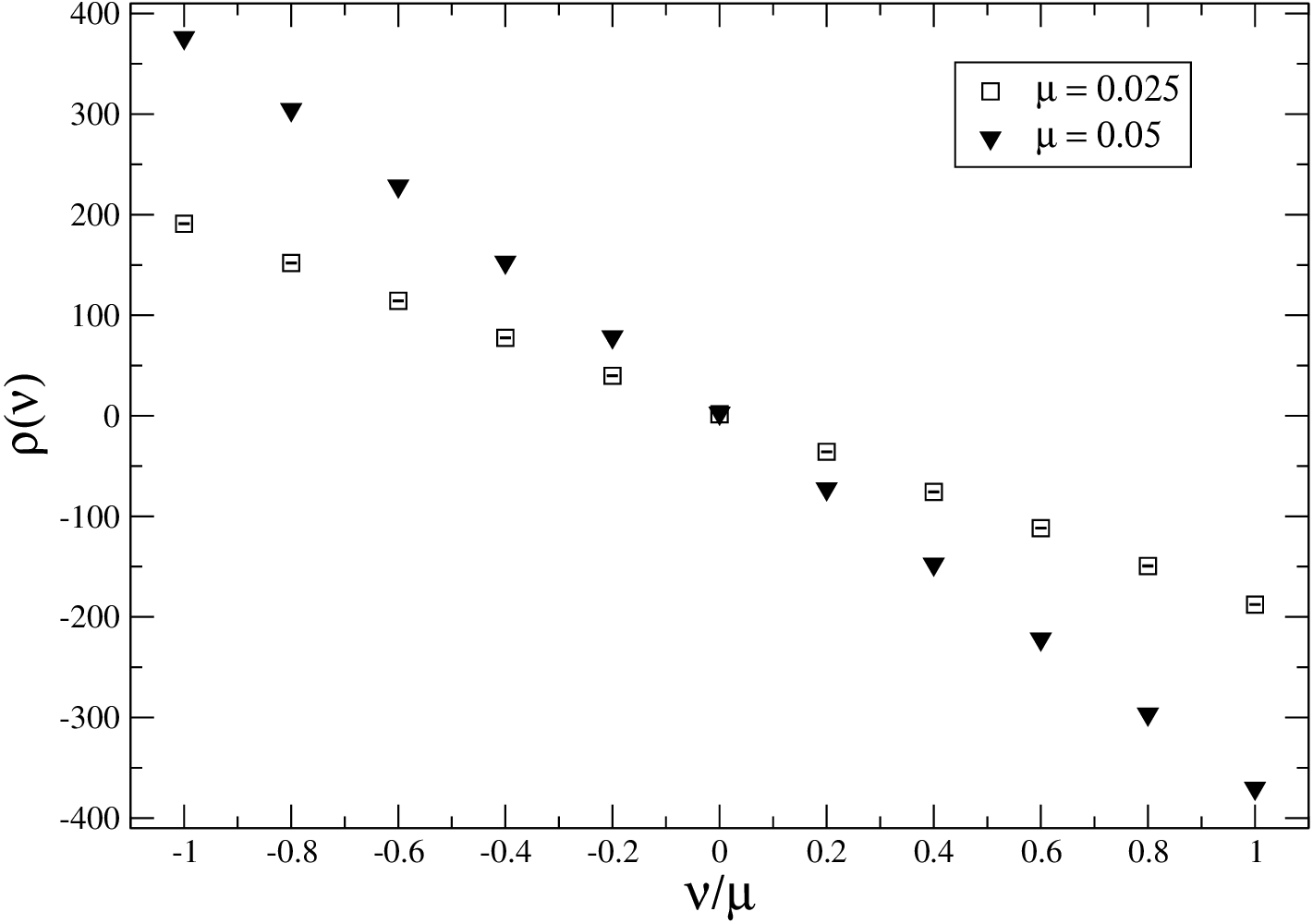}
\vspace{-0.cm}
\caption{
$\rho(\nu)$ for various values of $\mu$ at $\beta = 4.8$ 
and $L_s = 16$.
}
\label{fig3} 
\vspace{-0.cm}
\end{figure}

\begin{figure}[b!]
\includegraphics*[width=1.0\columnwidth]{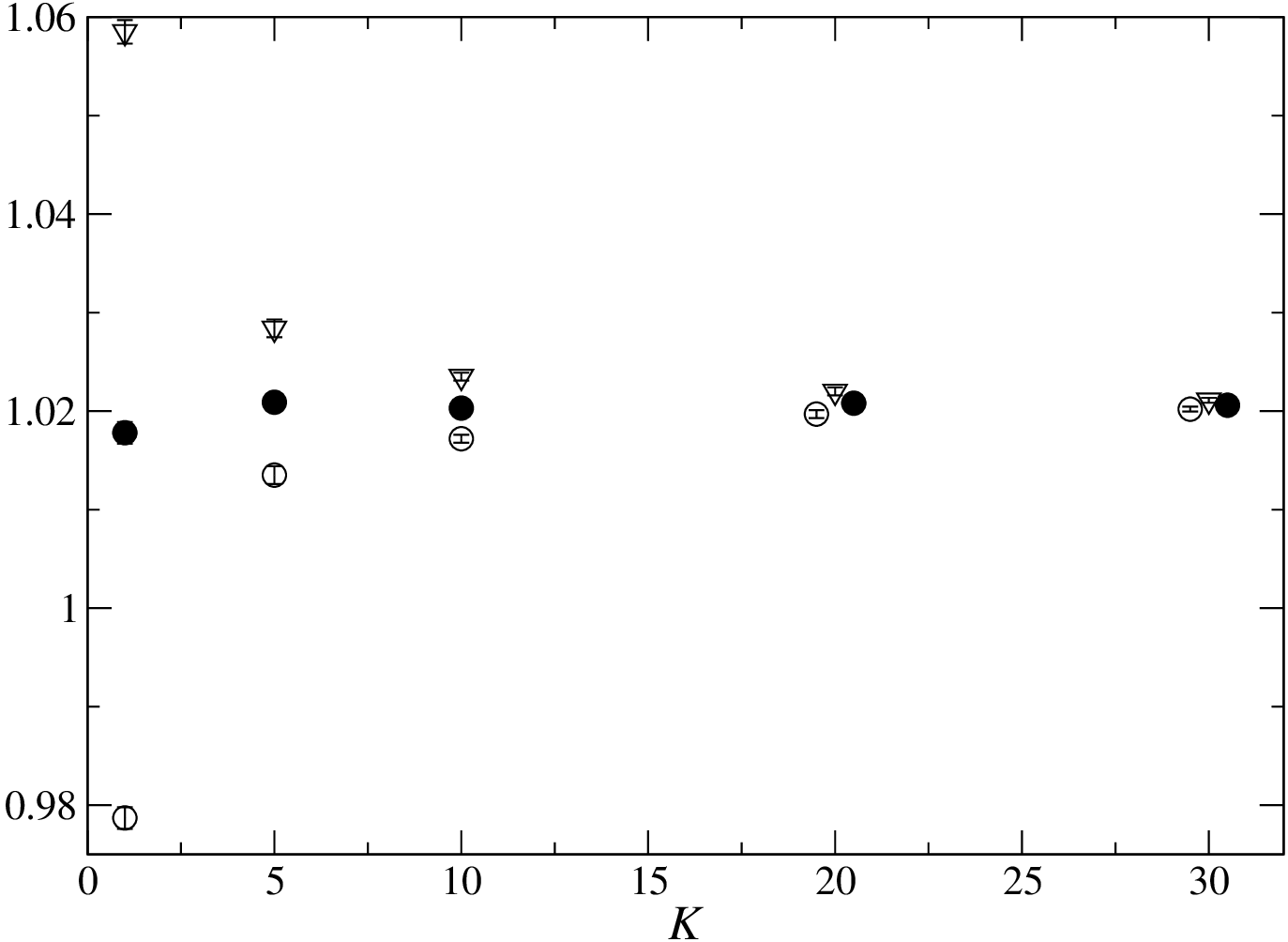}
\vspace{-0.cm}
\caption{
$r_k^+$ 
(blank triangles) and the inverse of $r_k^-$ 
(blank circles) defined in Eq.~(\ref{trick}), together with 
their geometric mean, i.e. $\sqrt{r_k}$ (filled circles).
Data are showed for 
$L_s = 4$, $\mu = 0.05$, $\nu = -0.045$
and $\delta \nu = 0.005$.
}
\label{fig:noise} 
\vspace{-0.cm}
\end{figure}

The second method is not conceptually different
from the first one: finite free energy differences are 
computed in this case instead of derivatives.
However the numerical procedure is different and it is not clear
apriori which approach is more convenient.
In the second case no numerical integration must be performed,
however one has the drawback that the exponential of a noisy
unbiased estimator is biased, hence a large number $K$ of vectors
must be used  and the final result must be checked to be independent
of $K$. Moreover the systematic error involved in the truncation
of the logarithm expansion, Eq.~(\ref{logexp}), must be properly
estimated and kept under control.

\section{Numerical results}
\label{results}

We have tested our methods for the theory with 8 staggered flavors
of mass $a m = 0.1$. We will present results
obtained on $L_s^3 \times L_t$ lattices with $L_t = 4$ and
$L_s = 4,8,16$. At zero chemical potential this theory presents
a strong first order deconfinement/chiral transition, the critical
coupling being $\beta_c \sim 4.7$ for $L_t = 4$. We have performed
simulations both in the deconfined region ($\beta = 4.8$) 
and in the low temperature confined region ($\beta = 4.6$).
On the smallest lattice ($L_s = 4$) we will compare our results
directly with those obtained at real isospin chemical potential
by a direct evaluation of the determinant phase. 
Numerical simulations have been performed mostly on the APEmille facility
in Pisa. The INFN apeNEXT facility in Rome has been used
for the results on the largest lattice. 
The standard exact HMC algorithm~\cite{Gottlieb:1987mq}
 has been used with trajectories
of length 1.

A collection of the results obtained for
imaginary chemical potentials
is reported in Table~\ref{table1}.

\subsection{Systematic errors and comparison of the methods}

In Figs.~\ref{fig1}, \ref{fig2} and \ref{fig3} we report various determinations
of $\rho(\nu)$ (minus the imaginary part of the baryon
number, Eq.~(\ref{rhodef})) obtained on discrete sets of points. 

It is apparent that $\rho(\nu)$ is a very smooth function of $\nu$ 
in all explored cases and independently of the lattice size
and of the explored phase (confined or deconfined)\footnote{
Clearly one expects a non-smooth behaviour if 
$Z(i \mu, i \mu)$ and $Z_0 \equiv Z(i \mu, - i \mu)$ belong to 
two different phases, so that some transition is met when
$\nu$ goes from $ - \nu$ to $ \nu$, however in these cases
analytic continuation itself is not applicable.
}. 
In most
cases it can even be approximated by a linear function; therefore numerical 
integration turns out to be an easy task. We have adopted a simple
linear interpolation between consecutive points to obtain the results
given in Table~\ref{table1}, the reported errors derive from 
standard error propagation of the statistical errors of the single
data points. We have verified, by changing the order of the
interpolating polynomial, that the systematic error related to
the interpolation-integration procedure is negligible with respect
to the statistical one.

Concerning the second method described in Section II, we have adopted
a standard trick~\cite{deforc} in order to reduce systematic effects.
Each partial ratio $r_k$ in Eq.~(\ref{2ndmet}) has been rewritten
as
\beq
\hspace{-25pt}
&r_k& \hspace{-5pt}
 = \left\langle
{\det M (i ( \nu + \delta \nu)) \over \det M (i \nu)}
\right\rangle_{i \mu, i \nu} = {r_k^+ \over r_k^-}
\nonumber \\
\hspace{-25pt}
& & \nonumber \\
\hspace{-25pt}
&\equiv& \hspace{-5pt}
{\left\langle
{\det M (i ( \nu + \delta \nu)) / \det M (i (\nu + {\delta \nu \over 2}))}
\right\rangle_{i \mu, i (\nu + {\delta \nu \over 2})}
\over
\left\langle
{\det M (i (\nu)) / \det M (i (\nu + {\delta \nu \over 2}))}
\right\rangle_{i \mu, i (\nu + {\delta \nu \over 2})}} \, .
\label{trick}
\eeq
$r_k$ can again be evaluated in a single 
simulation and a jackknife
analysis has to be applied to obtain a correct error estimate.
Two major benefits derive
in this case. First, the reduced value $\delta \nu /2$ greatly
improves the convergence of the logarithm expansion in 
Eq.~(\ref{logexp}). Second, the bias introduced by the finite
number of noisy estimators, see Eqs.~(\ref{logexp}) and 
(\ref{noise}), gets largely cancelled in the ratio. 
That is apparent from Fig.~\ref{fig:noise}, where we plot
$r_k^+$ and the inverse of $r_k^-$ defined in Eq.~(\ref{trick}),
and
their geometric mean (i.e. $\sqrt{r_k}$), as a 
function of the number $K$ of noise vectors, in one particular
sample case. It is clear that, while the single factors
have a relatively slow convergence, their product is stable
from $K = 5$ on. We have however always used $K = 30$ 
in our determinations.

\begin{table}
\begin{center}
\begin{tabular}{|r|c|l|l|l|r|}
\hline $L_s$ & $\beta$ & ${\rm Im}(\mu)$ &  method & $\langle e^{i 2 \theta}
\rangle_{i \mu} $ & HMC trajs \\
\hline
\hline  4 &  4.8  & 0.025  & DER(10) & 1.00322(42) & 700k   \\
\hline  4 &  4.8  & 0.025  & RAT(5)  & 1.0030(18)  & 150k   \\
\hline  4 &  4.8  & 0.025  & RAT(10) & 1.0028(11)  & 300k   \\
\hline  4 &  4.8  & 0.025  & direct  & 1.0033(11)  &  40k   \\
\hline  4 &  4.8  & 0.05   & DER(20) & 1.0108(11)  & 800k   \\
\hline  4 &  4.8  & 0.05   & RAT(10) & 1.0122(16)  & 500k   \\
\hline  4 &  4.8  & 0.075  & DER(15) & 1.0266(17)  & 350k   \\
\hline  4 &  4.8  & 0.10   & DER(20) & 1.0454(16)  & 700k   \\
\hline  4 &  4.8  & 0.20   & DER(16) & 1.283(8)    & 700k   \\
\hline  8 &  4.8  & 0.025  & DER(10) & 1.0164(19)  & 150k   \\
\hline  8 &  4.8  & 0.025  & RAT(5)  & 1.0200(50)  &  50k   \\
\hline 16 &  4.8  & 0.025  & DER(10) & 1.0732(85)  &  60k   \\
\hline 16 &  4.8  & 0.025  & RAT(5)  & 1.053(33)   &  40k   \\
\hline 16 &  4.8  & 0.05   & DER(10)  & 1.368(30)   &  40k  \\
\hline  4 &  4.6  & 0.025  & DER(5)  & 1.0061(10)  & 200k   \\
\hline  4 &  4.6  & 0.05   & DER(10) & 1.0270(15)  & 350k   \\
\hline
\end{tabular}
\end{center}
\caption{\label{table1}
Collection of determinations of the average phase factor continued
to imaginary values of $\mu$ for various parameter sets and
computation methods. In the fourth column the method used to obtain
the determination is described: DER(N)
stands for the integration of the first derivative $\rho$ determined
on a discrete set of (N+1) points, Eq.~(\ref{1stmet}); RAT(N) stands for the evaluation
of N intermediate ratios $r_k$, Eq.~(\ref{2ndmet}). Finally on the
smallest lattices also a direct determination of the expectation
value
in Eq.~(\ref{phase2}) is reported for comparison.
}
\end{table}

Regarding the logarithm expansion, Eq.~(\ref{logexp}), we have
always adopted a third order approximation: in all cases
the discrepancy with the result obtained at the second order is
at least one order of magnitude smaller than the statistical
uncertainty. The fact that the systematic error related to this 
expansion is well under control can be also appreciated from
Table~\ref{table1}, second and third row, showing that 
the determination of $\langle e^{i 2 \theta}
\rangle_{i \mu} $ is stable against the variation of the 
number of intermediate ratios.

Let us now come to the comparison between the two methods.
While they always give perfectly compatible results, thus 
confirming the absence of appreciable systematics, it is clear
from Table~\ref{table1} that, with a comparable numerical effort
(in the last column we give the total number of Monte-Carlo 
 trajectories used for each determination), the method
described by Eq.~(\ref{1stmet}) (integration of the derivative) 
furnishes more accurate determinations. We have therefore chosen
this method in order to perform more extensive 
studies of $\langle e^{i 2 \theta}
\rangle_{i \mu} $.

\begin{figure}[t!]
\includegraphics*[width=1.0\columnwidth]{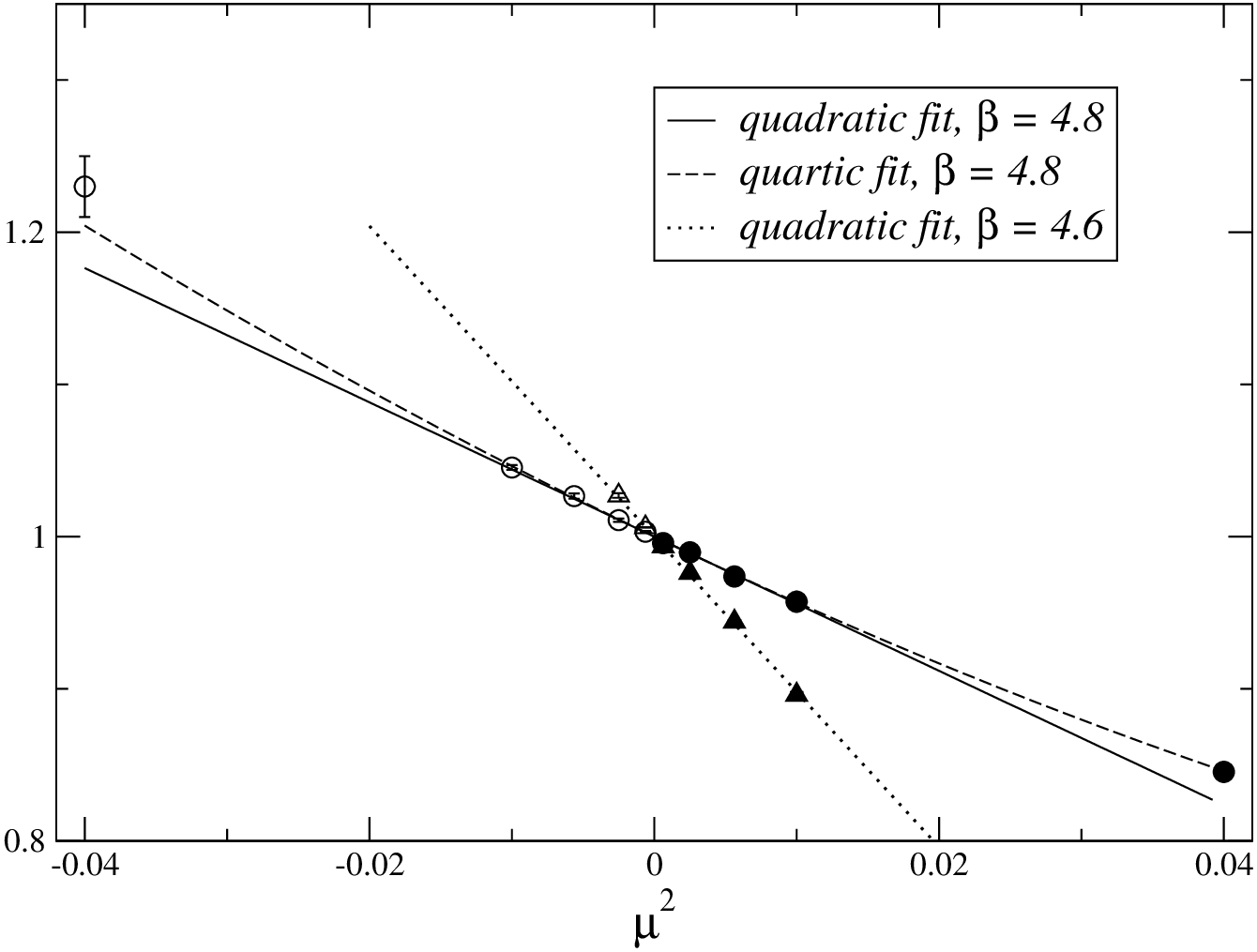}
\vspace{-0.cm}
\caption{
$\langle e^{i 2 \theta}
\rangle$ computed for different values of $\mu^2$ 
at $\beta = 4.8$ and $\beta = 4.6$ on a $4^4$ lattice.
Best fit quadratic and quartic functions in $\mu^2$ are
displayed, showing good validity of analytic continuation.
}
\label{fig:ana} 
\vspace{-0.cm}
\end{figure}

\begin{figure}[b!]
\includegraphics*[width=1.0\columnwidth]{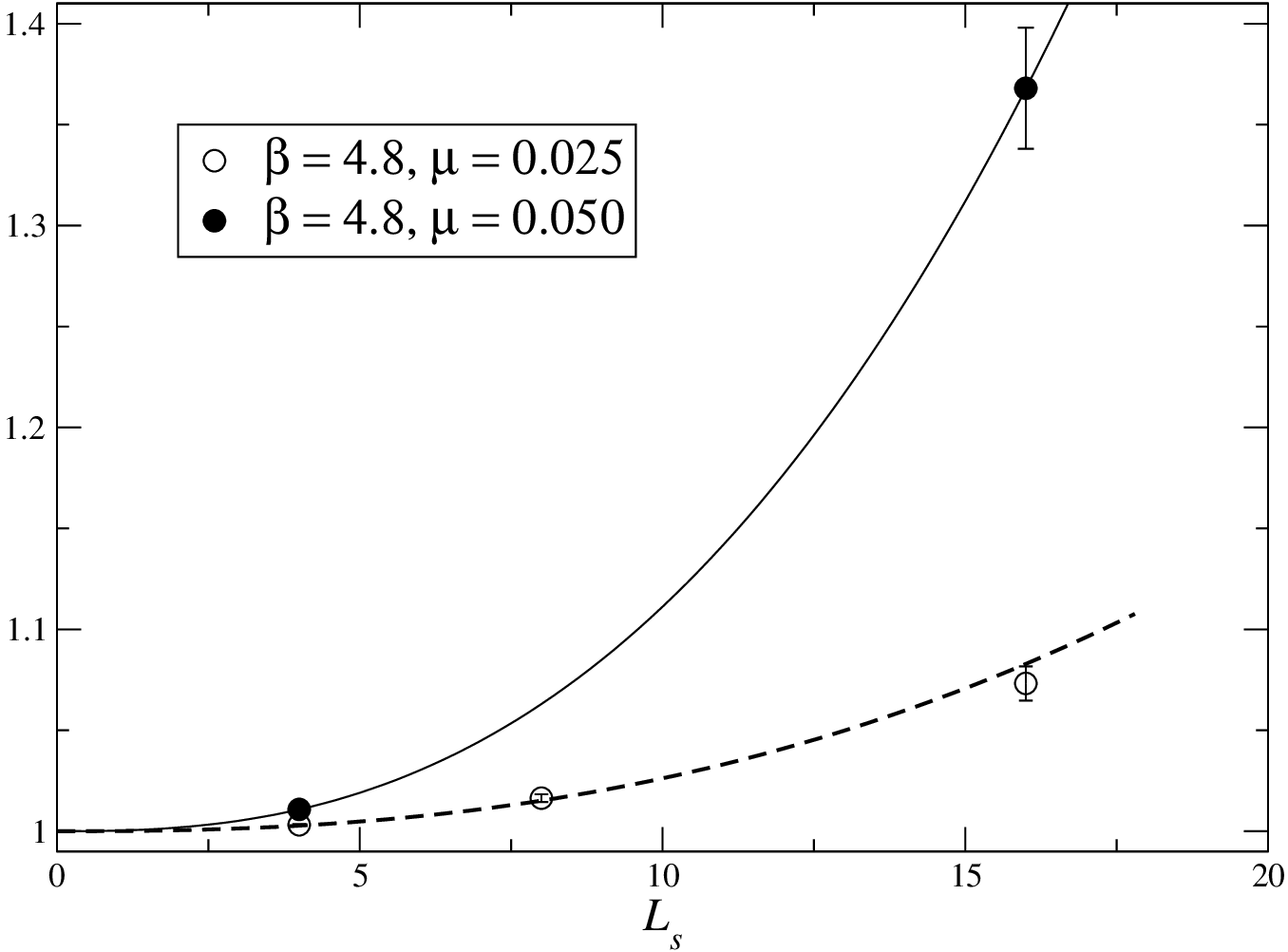}
\vspace{-0.cm}
\caption{
$\langle e^{i 2 \theta} \rangle_{i \mu}$ as a function of the 
spatial lattice size $L_s$ for two values of $i \mu$. A best fit 
according to Eq.~(\ref{fitvol}) is reported in both cases.
}
\label{fig:vol} 
\end{figure}

\subsection{Test of analytic continuation}

The average phase factor
computed at finite isospin chemical potential, at variance
with that computed in the quenched theory, is expected~\cite{splitt1,splitt2}
 to be an
analytic function of $\mu^2$
around $\mu^2 = 0$\footnote{
It is even in $\mu$ for symmetry reasons.}. 
We can test directly analytic
continuation by comparing our results with 
direct determinations of $\langle e^{i 2 \theta}
\rangle_{i \mu} $ performed at real chemical potentials:
this is done only for the smallest lattice ($L_s = 4$), where
the second determination is easily affordable.

We plot in Fig.~\ref{fig:ana} results obtained at $\beta = 4.6$
and $\beta = 4.8$. The whole set of results 
obtained at real chemical potentials ($\mu^2 > 0$) and 
imaginary chemical potentials ($\mu^2 < 0$) can be described
by a simple quadratic behaviour
\beq
\langle e^{i 2 \theta} \rangle = 1 + A \mu^2
\eeq
in a range $|\mu^2| \leq 0.01$, with
$A = -4.41(9)$ and $\chi^2/{\rm d.o.f.} \simeq 1.5$ for $\beta = 4.8$ 
and 
$A = -10.2(3) $ and $\chi^2/{\rm d.o.f.} \simeq 1.8$ for $\beta = 4.6$.
If the range of values is extended a quartic term is necessary
\beq
\langle e^{i 2 \theta} \rangle = 1 + A \mu^2 + B \mu^4
\eeq
as shown in the figure. We obtain, at $\beta = 4.8$,
$A = -4.48(8)$, $B = 15.7 \pm 2.5$  and $\chi^2/{\rm d.o.f.} \simeq 1.3$.

Analyticity around $\mu^2 = 0$ is therefore well verified.
We stress 
that at $\beta = 4.8$ our largest value of the imaginary chemical 
potential is still below the first Roberge-Weiss phase transition
at ${\rm Im} (\mu) = \pi / (3 L_t)$, hence within the expected
range of validity of analytic continuation for $\mu^2 < 0$
at high temperatures.

\subsection{Large volume scaling}

We have performed numerical simulations at different values
of $L_s$ in order to test both the behaviour of
$\langle e^{i 2 \theta} \rangle $ and the efficiency of our 
method as the lattice volume is increased.

In Fig.~\ref{fig:vol} we report determinations performed
at fixed values of $i \mu$ and variable $L_s$ at $\beta = 4.8$.
A behaviour 
\beq
\langle e^{i 2 \theta} \rangle = 1 + C L_s^\gamma
\label{fitvol}
\eeq
well describes the data with $\gamma \sim 2.5$ for both
values of $i \mu$.

Concerning the numerical efficiency, we notice that 
to obtain comparable uncertainties (of the order of 10~\%)
for $\langle e^{i 2 \theta} \rangle - 1$, on the largest
lattice ($16^3 \times 4$) we needed a CPU time which is
less than one order of magnitude bigger than what 
needed on the smallest lattice ($4^4$). Considering
that the two lattice volumes differ by a factor 64, we deduce
that, at least for the quark mass considered in the present
study, our method requires a numerical effort which scales
in an affordable way with the lattice size.

\section{Conclusions} 
\label{conclusions}

We have presented two different techniques,
described respectively by Eq.~(\ref{1stmet}) and 
Eq.~(\ref{2ndmet}), for an efficient numerical
determination of the average phase factor of the fermionic determinant
continued to imaginary values of the chemical potential.
We have applied both methods to QCD with 8 dynamical staggered
flavors, verifying the absence of uncontrolled systematic 
effects and performing a comparison of the efficiencies, 
with the conclusion that the method 
based on the integration of the imaginary part of the 
baryon density, Eq.~(\ref{1stmet}), is numerically more convenient.
A fair good scaling of the efficiency is observed as the 
lattice volume is increased.
We have also directly tested, on small lattices, the analiticity 
of the average phase factor around $\mu^2 = 0$.
The method proposed and tested in the present paper will be used in 
the future to perform more extensive studies, with more physical
quark masses and
number of flavors, of the average
phase factor continued to imaginary chemical potentials.

\end{document}